\documentclass[twocolumn,showpacs,preprintnumbers,amsmath,amssymb]{revtex4}

\usepackage{graphicx}
\usepackage{dcolumn}
\usepackage{bm}
\def\i{i}
\def\d{d}
\def\e{e}
\def\vector#1{{\boldsymbol{#1}}}

\def\vA{{\vector A}}

\def\vH{{\vector H}}

\def\vp{{\vector p}}
\def\vq{{\vector q}}

\def\vr{{\vector r}}

\def\vv{{\vector v}}

\def\vPi{{\vector \Pi}}

\def\dps{\displaystyle}

\def\hsp#1{\hspace{#1ex}}

\def\lsim{\stackrel{{\textstyle<}}{\raisebox{-.75ex}{$\sim$}}}
\def\gsim{\stackrel{{\textstyle>}}{\raisebox{-.75ex}{$\sim$}}}

\def\eq.#1{Eq.~(\ref{#1})}
\def\eqs.#1{Eqs.~(\ref{#1})}
\def\refeq#1{(\ref{#1})}

\def\mue{\mu_e}

\newcommand\Equation[2]{
\begin{equation}\label{#1}
#2
\end{equation}
}

\begin{document}
\preprint{APS/123-QED}

\title{
Crossover from the vortex state 
to the Fulde--Ferrell--Larkin--Ovchinnikov state 
in quasi-two-dimensional superconductors 
}


\author{Hiroshi Shimahara}



\affiliation{
Department of Quantum Matter Science, ADSM, Hiroshima University, 
Higashi-Hiroshima 739-8530, Japan
}


\date{\today}

\begin{abstract}
We examine the coexistence of the vortex state and 
the Fulde--Ferrell--Larkin--Ovchinnikov (FFLO) state 
in quasi-two-dimensional type-II superconductors 
and the crossover from the coexistence state to the pure FFLO state 
when the Maki parameter $\alpha$ increases. 
The pure FFLO state, characterized by finite center-of-mass momenta 
$\vq \ne {\vector 0}$ of Cooper pairs 
occurs in the two-dimensional limit, 
when the magnetic field is parallel to the conductive plane. 
The vectors $\vq$ are determined 
from the Fermi-surface structure and pairing anisotropy, 
and become finite below a \mbox{temperature $T^{*}$}. 
In quasi-two-dimensions, 
because of the orbital pair-breaking effect, 
the coexistence state characterized by $(n,q_{\parallel})$ occurs, 
where $n$ and $q_{\parallel}$ denote 
the Landau level index of the vortex state and 
the wave number of the additional FFLO modulation along the magnetic field. 
We obtain the $\alpha$ dependence of the upper critical field 
by numerical calculations. 
The upper critical field exhibits a cascade curve 
in the $H$--$T$ phase diagram. 
It is analytically shown that $n$ diverges 
in the two-dimensional limit $\alpha \rightarrow \infty$ 
below $T^{*}$. 
In this limit, the upper critical field equation 
of the coexistence state is reduced to that of the FFLO state. 
A relation between $n$ of the coexistence state 
and $q_{\perp}$ of the pure FFLO state is obtained, 
where $q_{\perp}$ denotes the component of $\vq$ perpendicular to 
the magnetic field. 
It is found that the pure FFLO state is nothing but 
the vortex state with infinitely large $n$ 
as is known in two-dimensional superconductors 
in a tilted magnetic field. 
The vortex state with large $n$ can be regarded as the FFLO state 
with non-zero $q_{\perp}$ in three dimensions. 
\end{abstract}


\maketitle


\section{\label{sec:introduction}
INTRODUCTION
}

In type-II superconductors, an applied magnetic field 
destroys the superconductivity by two kinds of pair-breaking effects: 
the orbital magnetic and Pauli paramagnetic pair-breaking effects~\cite{Cha62}. 
We can define the pure orbital limit $H_{\rm c20}$ and 
the Pauli paramagnetic limit $H_{\rm P}$, 
which are theoretically obtained 
by taking into account only the orbital effect 
and paramagnetic effect, respectively. 
The Maki parameter $\alpha \equiv \sqrt{2} H_{{\rm c}20}/H_{\rm P}$ 
expresses the strength ratio of the two pair-breaking effects.

In conventional metal superconductors, 
the orbital pair-breaking effect dominates the system 
($H_{\rm c20} \ll H_{\rm P}$) 
because of the large Fermi velocity. 
Partial destruction of the superconductivity due to the orbital effect 
creates vortexes, 
which form a lattice below the upper critical field, 
and causes the order parameter to become nonuniform.

In contrast, in purely Pauli limited superconductors, 
another type of nonuniform superconductivity has been proposed 
by Fulde and Ferrell~\cite{Ful64} and Larkin and Ovchinnikov~\cite{Lar64}. 
In magnetic fields, 
the Fermi surfaces of the up- and down-spin electrons are displaced 
due to the Zeeman energy. 
If the up- and down-spin electrons on the displaced Fermi surfaces 
form Cooper pairs, 
they should have a finite center-of-mass momentum. 
The superconducting state of such Cooper pairs is called 
the Fulde--Ferrell--Larkin--Ovchinnikov (FFLO) state. 
It is easily verified that the finite center-of-mass momentum 
results in spatial modulations and nodes of the order parameter 
in real space. 
As a result, 
spin polarization energy is gained by the depaired electrons 
near the nodes, while the condensation energy is lost.

Therefore, 
a necessary condition for the occurrence of the FFLO state 
is that the superconductivity survives in high fields such that 
$\mue H \sim \Delta_0$, 
for which the loss of condensation energy due to modulation 
of the order parameter is compensated by a gain in polarization energy. 
Here, $\mue$ and $\Delta_0$ 
denote the electron magnetic moment 
and the zero field energy gap of the superconductivity. 
This condition can be expressed as $H_{\rm c2} \sim H_{\rm P}$, 
where $H_{\rm c2}$ denotes the upper critical field, 
because $H_{\rm P} \sim \Delta_0$. 
Therefore, the orbital pair-breaking effect needs to be very weak 
for the FFLO state to occur.

This is one of the reasons why the FFLO state has not been observed 
in conventional metal superconductors. 
Gruenberg and Gunther found that 
the FFLO state occurs only when the Maki parameter $\alpha$ is 
large ($\alpha \gsim 1.8$) in isotopic superconductors~\cite{Gru66}. 
Such a large Maki parameter is usually difficult to realize 
in alloy type II superconductors, 
though it is achievable in some exotic superconductors, 
such as organic superconductors~\cite{Shi08a}, 
heavy fermion superconductors~\cite{Mat07}, 
and oxide superconductors, 
because of their narrow electron bands, large effective masses, 
and quasi-two-dimensionality.

Because of the orbital pair-breaking effect, 
and the formation of the vortex lattice state, 
the dependence of the order parameter on the spatial coordinates 
perpendicular to the magnetic field 
is described by the superposition of the Abrikosov functions. 
Therefore, there may be additional modulations 
due to the FFLO state only in the direction parallel to the magnetic field, 
as Gruenberg and Gunther have proposed~\cite{Gru66}. 
We examined the coexistence of the vortex states 
with higher Landau level indexes $n$ and the FFLO state 
in $d$-wave superconductors in our previous paper~\cite{Shi96}, 
and obtained phase diagrams for some cases. 
Recently, in a model of the quasi-two-dimensional heavy fermion 
superconductor ${\rm CeCoIn_5}$, 
Adachi and Ikeda have shown 
that the first-order phase-transition 
to a coexistence state occurs 
taking into account the higher Landau level indexes~\cite{Ada03}.

Bulaevskii examined film superconductors at $T = 0$ 
in tilted magnetic fields, 
and obtained cascade transitions between the states with different $n$'s 
when the direction of the magnetic field changes. 
It was shown that the upper critical field tends to approach 
that of the FFLO state when the magnetic field 
approaches the parallel direction. 
Buzdin and Brison also obtained cascade transitions 
at finite temperatures in two- and three-dimensional 
isotropic superconductors~\cite{Buz96}. 
We extended Bulaevskii's theory to finite temperatures 
in $s$- and $d$-wave superconductors~\cite{Shi97a}. 
We also reproduced the cascade transitions solving the gap equation, 
and clarified the behavior in the limit of a parallel field. 
We analytically demonstrated that 
the Landau level index $n$ of the vortex state 
diverges in the temperatures region where the FFLO state occurs in the limit. 
As a result, the envelope of the cascade transition lines 
approaches the FFLO critical field when the field orientation 
approaches the parallel direction. 
The coexistence state is continuously reduced to the pure FFLO state. 
We obtained a relation between $n$ and the FFLO vector $\vq$, 
which connects the vortex states and the FFLO state in the limit. 
The present study extends our previous theory 
to the coexistence state in three dimensions.

Near critical fields, 
the order parameter of the pure FFLO state can be expressed 
by a linear combination of exponential functions 
$\exp[ \i \vq_m \cdot \vr ]$, where $\vq_m$ are 
the degenerate FFLO wave vectors 
with optimum values determined from 
the structures of the Fermi surfaces and the pairing interactions, 
and the temperature. 
The free energies of such states were compared 
in a three dimensional isotropic system 
by Larkin and Ovchinnikov~\cite{Lar64}, 
Matsuo {\it et al.}~\cite{Mat98}, 
Bowers and Rajagopal~\cite{Bow02}, 
and Mora and Combescot~\cite{Mor05}, 
and in two-dimensional systems 
by the author~\cite{Shi98a} 
and Mora and Combescot~\cite{Mor04}. 
Larkin and Ovchinnikov~\cite{Lar64} showed that the state expressed by 
$\Delta(\vr) \sim \cos (\vq \cdot \vr)$ has a lower free energy 
than the state $\Delta(\vr) \sim \exp [\i \vq \cdot \vr]$ 
proposed by Fulde and Ferrell. 
The author has shown that 
the square, triangular, and hexagonal states have lower free energies 
than the state $\Delta(\vr) \sim \cos (\vq \cdot \vr)$ 
at low temperatures in two dimensions~\cite{Shi98a}. 
Mora and Combescot have shown that 
the states described many cosine functions have lower free energies 
than the state $\Delta(\vr) \sim \cos (\vq \cdot \vr)$ 
in three dimensions when the first-order phase transition 
is taken into account~\cite{Mor05}. 
The possibility of the first-order transition was addressed 
also by Larkin and Ovchinnikov~\cite{Lar64}.

When the magnetic field is not oriented in the optimum direction of $\vq$ 
of the pure FFLO state, 
or when more than two $\vq_m$'s contribute to the linear combination 
for the pure FFLO state, 
it may appear that 
the coexistence state is not reduced to the pure FFLO state 
in the limit $\alpha \rightarrow \infty$, 
because in the coexistence state $\vq$ can have a nonzero component 
only in the direction of the magnetic field. 
In actuality, however, the Landau level index $n$ of the coexistence 
state diverges and the components of $\vq$ perpendicular to the 
magnetic field are realized~\cite{Shi08}. 
This behavior is analogous to that 
in two-dimensional systems in a tilted magnetic field. 
In the present paper, we demonstrate this behavior 
in quasi-two-dimensional systems 
by an analytical proof and concrete numerical calculations.

For simplicity, we adopt an effective mass model 
assuming that the first-order transition is suppressed, 
to demonstrate the continuity between the coexistence states and 
the pure FFLO state. 
In actuality, in the effective mass model with very large $\alpha$, 
the first-order transition would occur at 
a slightly higher field than 
that of the second-order transition~\cite{Mor05}. 
In this case, the second-order transition curve below the first-order 
transition curve is regarded as that of the metastable transition, 
which can be realized when the system is supercooled.

Recently, we have obtained a result which also suggests that 
the Landau level index $n$ increases with $\alpha$ 
in an anisotropic Ginzburg--Landau model 
near the tricritical point~\cite{Den09}. 
Comparing the phase diagrams with and without the orbital effect, 
we have found that the areas of the coexistence states with $n > 0$ 
in the former phase diagrams correspond to 
the areas of the pure FFLO state with $q_{\perp} \ne 0$ 
in the latter phase diagrams.

In section 2, we present our formulation. 
In sections 3 and 4, we examine the pure FFLO state 
and the coexistence state, respectively. 
In section 4, we show the crossover from the pure FFLO state 
to the coexistence state and the limit $\alpha \rightarrow \infty$. 
In section 5, we show the numerical result for finite temperatures. 
In section 6, we summarize and discuss the results.

\section{\label{sec:formulation}
FORMULATION
}

We examine a model described by the Hamiltonian 
\Equation{eq:H}
{
     H = H_0 + H_m + H' , 
     }
with 
\Equation{eq:H0}
{
     H_0 = \sum_{\mu \sigma} 
          \int \! \d^3 \vr \, 
          \psi_{\sigma}^{\dagger}(\vr) 
           \frac{1}{2 m_{\mu}} 
                \bigl ( - \i \hbar \frac{\partial}{\partial x_{\mu}} 
                        - \frac{e}{c} A_{\mu} 
                \bigr )^2 
          \psi_{\sigma}(\vr) , 
     }
\Equation{eq:Hm}
{
     H_m = \sum_{\sigma} 
          \int \! \d^3 \vr \, 
          \sigma h \, \psi_{\sigma}^{\dagger}(\vr) \psi_{\sigma}(\vr) , 
     }
\Equation{eq:Hint}
{
     H' = \int \! \d^3 \vr 
          \int \! \d^3 \vr' \, 
          \psi_{\uparrow}^{\dagger} (\vr) \psi_{\uparrow}(\vr)
            V(\vr - \vr') 
          \psi_{\downarrow}^{\dagger} (\vr') \psi_{\downarrow}(\vr') . 
     }
Here, we have defined the effective masses 
$m_1 = m_x$, $m_2 = m_y$, $m_3 = m_z$, 
Zeeman field $h = \mue |\vH|$, and vector potential $\vA$, 
where $\mue$ and $\vH$ denote 
the magnitude of the electron magnetic moment 
and the magnetic field $\vH = {\rm rot} \vA$. 
We consider pairing interactions of the form 
\Equation{eq:Valpha}
{
     V(\vp,\vp') = g_{\alpha} \gamma_{\alpha}({\hat \vp}) 
                              \gamma_{\alpha}({\hat \vp}') , 
     }
where the suffix $\alpha$ expresses a symmetry.

In the effective mass model of Eq.~\refeq{eq:H0}, 
it is convenient to define 
${\tilde \vr} = ({\tilde x}_1,{\tilde x}_2,{\tilde x}_3)$ 
by a scale transformation 
\Equation{eq:scale}
{
     \sqrt{m_{\mu}} \, x_{\mu} \equiv \sqrt{\tilde m} \, {\tilde x}_{\mu}
     }
with ${\tilde m} = (m_x m_y m_z)^{1/3}$. 
Then, Eq.~\refeq{eq:H0} is written as 
\Equation{eq:H0scaled}
{
     H_0 = \sum_{\mu \sigma} 
          \int \! \d^3 \vr \, 
          \psi_{\sigma}^{\dagger}(\vr) 
           \frac{1}{2 {\tilde m}} 
                \bigl ( - \i \hbar \frac{\partial}{\partial {\tilde x}_{\mu}} 
                        - \frac{e}{c} {\tilde A}_{\mu} 
                \bigr )^2 
          \psi_{\sigma}(\vr) , 
     }
where we have defined 
\Equation{eq:Ascale}
{
     {\tilde A}_{\mu} \equiv \sqrt{\frac{\tilde m}{m_{\mu}}} A_{\mu} . 
     } 
We also define ${\tilde \vp} = ({\tilde p}_1,{\tilde p}_2,{\tilde p}_3)$ 
with 
${\tilde p}_{\mu} = (\tilde m/m_{\mu})^{1/2} p_{\mu}$, 
so that $\vr \cdot \vp = {\tilde \vr} \cdot {\tilde \vp}$. 
Then, the Fermi surface in ${\tilde \vp}$ space becomes spherically symmetric, 
and we can define a constant Fermi momentum ${\tilde p}_F$ 
and Fermi velocity 
${\tilde \vv}_F = {\tilde v}_F {\tilde \vp}/|{\tilde \vp}|$ 
with a constant magnitude 
${\tilde v}_F = {\tilde p}_F/{\tilde m}$, 
for the scaled momentum ${\tilde \vp}$.

Now, we derive the gap equation. 
The calculation is a straightforward extension of 
the previous studies~\cite{Luk87,Shi96,Shi97a,Sug06}. 
Near the second-order phase transition, the gap function has a form 
\Equation{eq:gapfunction}
{
     \Delta(\vr,\vp) = \Delta_{\alpha}(\vr) \gamma_{\alpha}(\vp) , 
     }
and the gap equation is linearized as 
\Equation{eq:Hc2eq}
{
     \begin{array}{rcl}
     \lefteqn{ \dps{ 
     - \log \bigl ( \frac{T}{T_c^{(0)}} \bigr ) 
     \Delta_{\alpha}(\vr) 
       } } \\[12pt]
     & = & 
     \dps{ 
     \pi T \int_0^{\infty} \d t \frac{1}{\sinh( \pi T t )} 
          \int \frac{\d \Omega_{{\tilde \vp}'}}{4 \pi} 
          \bigl [ \gamma_{\alpha}( {\hat \vp}' ) \bigr ]^2 
         } \\[12pt]
     && 
     \dps{ 
          \times 
          \Bigl [ 1 - \cos \bigl [ t \bigl \{ 
               h - \frac{1}{2} {\tilde \vv'_F} \cdot {\tilde \vPi} 
          \bigr \} \bigr ] \Bigr ] 
          \Delta_{\alpha}(\vr) , 
            }
     \end{array}
     }
where $\vv'_F = {\tilde v}_F {\tilde \vp}'/|{\tilde \vp}'|$ 
and ${\tilde \vPi} = ({\tilde \Pi}_1, {\tilde \Pi}_2, {\tilde \Pi}_3)$ 
with 
\Equation{eq:Pidef}
{
     {\tilde \Pi}_{\mu} 
     = - \i \hbar \frac{\partial}{\partial {\tilde x}_{\mu}} 
                      - \frac{2 e}{c} {\tilde A}_{\mu} . 
     }
The upper critical field $H_{\rm c2}$ 
is the highest $|\vH|$ among those which give 
a nontrivial solution of $\Delta(\vr)$. 
We note that Eq.~\refeq{eq:Hc2eq} is the same as that of a system with 
a spherically symmetric Fermi surface except that 
the argument ${\hat \vp}' = \vp'/|\vp'|$ of 
$\gamma_{\alpha}( {\hat \vp}' )$ is different from the integral variable 
${\hat {\tilde \vp}}' 
= {\tilde \vp}'/|{\tilde \vp}'| = {\tilde \vp}'/{\tilde p}_F$. 
Therefore, when $\gamma_{\alpha}( {\hat \vp}' )$ is constant, 
it is easily verified 
that the mass anisotropy does not affect the upper critical field equation 
except that the vector potential is scaled 
as described in Eq.~\refeq{eq:Ascale}, 
since ${\hat {\tilde \vp}}'$ is only an integral variable. 
In contrast, for anisotropic superconductors, 
the mass anisotropy affects the upper critical field equation 
through the deformation of $\gamma_{\alpha}({\hat \vp}')$ 
when it is expressed in ${\tilde \vp}$ space~\cite{Sug06}.

\section{\label{sec:pureFFLOstate}
the pure FFLO state 
}

In this section, 
we briefly review the case in which the orbital pair-breaking effect 
is negligible. 
In this case, 
we can set $\vA = {\vector 0}$ in Eqs.~\refeq{eq:H0} and~\refeq{eq:Pidef}. 
Equation~\refeq{eq:Hc2eq} has a solution of the form 
\Equation{eq:FFstate}
{
     \Delta(\vr) \propto \exp[ \i {\tilde \vq} \cdot {\tilde \vr}/\hbar ] , 
     }
and is reduced to 
\Equation{eq:A0fromthefirst}
{
     \begin{array}{rcl}
     \lefteqn{ \dps{ 
     - \log \bigl ( \frac{T}{T_c^{(0)}} \bigr ) 
     = 
     \pi T \int_0^{\infty} \d t \frac{1}{\sinh( \pi T t )} 
       } } \\[12pt]
     && 
     \dps{ 
          \times 
          \int \frac{\d \Omega_{{\tilde \vp}'}}{4 \pi} 
          \bigl [ \gamma_{\alpha}( {\hat \vp}' ) \bigr ]^2 
          \Bigl [ 1 - \cos \bigl [ t \bigl \{ 
               h - \frac{1}{2} {\tilde \vv'_F} \cdot {\tilde \vq} 
          \bigr \} \bigr ] \Bigr ] . 
            }
     \end{array}
     }

For example, for $s$-wave pairing, $\gamma_s(\vp) = 1$, 
there is infinite degeneracy 
with respect to the direction of ${\tilde \vq}$, 
although the magnitude $|\tilde \vq|$ is uniquely determined 
so that the critical field is maximized. 
For non-$s$-wave pairing, 
both the direction and the magnitude of ${\tilde \vq}$ are optimized. 
Depending on the symmetries of $\gamma_{\alpha}(\vp)$ and the Fermi surface, 
and the temperature, 
there may be $2, 4, 8, 16 \cdots$-fold degeneracies 
with respect to the direction of ${\tilde \vq}$. 
We write the optimum ${\tilde \vq}$'s as ${\tilde \vq}_m$ 
with $m = 1,2,\cdots, M$. 
Below and near the upper critical field, 
the order parameter is expressed by a linear combination: 
\Equation{eq:linearcomb}
{
     \Delta(\vr) 
     = \sum_{m} \Delta_m \e^{\i {\tilde \vq}_m \cdot {\tilde \vr}} . 
     }
Among the states of this form, the physical state is 
that with the lowest free energy. 
Every degenerate ${\tilde \vq}_m$ does not necessarily appear in 
the linear combination of the physical state. 
The most well-known form is that expressed by the linear combination of 
$\e^{\i {\tilde \vq} \cdot {\tilde \vr}}$ 
and $\e^{- \i {\tilde \vq} \cdot {\tilde \vr}}$, {\it i.e.}, 
$\Delta(\vr) \propto \cos ({\tilde \vq} \cdot {\tilde \vr})$. 
For the second-order transition and $s$-wave pairing, 
this state is the physical state in the effective mass model.

Here, we note that ${\tilde \vq}_m$ are not necessarily 
parallel to the magnetic field, 
when the orbital effect is negligible. 
The number and directions of the optimum ${\tilde \vq}_m$ 
which contribute to the physical state also depend on 
the structures of the Fermi surface 
and the pairing interactions~\cite{Shi94,Shi97b}, 
and the temperature~\cite{Mak96,Shi97a,Yan98,Shi02}. 
In the present effective mass model, 
since the electron dispersion becomes isotropic in ${\tilde \vp}$ space 
as seen in Eq.~\refeq{eq:A0fromthefirst}, 
the Fermi-surface anisotropy does not remove 
the infinite degeneracy of the optimum ${\tilde \vq}_m$ 
for $s$-wave pairing.

\section{\label{sec:coexistencestate}
coexistence state 
}

In this section, we take into account both the orbital and paramagnetic 
pair-breaking effects. 
After deriving the upper critical field equation, 
we take the limit of a weak orbital effect.

For a magnetic field $\vH = (0,0,H)$, 
we define $\vA = ( - H y, 0, 0 )$ with an appropriate gauge. 
The scale transformation described above gives 
${\tilde A}_x 
= - {\tilde H}{\tilde y}$ with 
${\tilde H} = ({\tilde m}/\sqrt{m_x m_y}) H$. 
We define boson operators by 
\Equation{eq:etatilde}
{
     \begin{array}{rcl}
     \tilde \eta 
          & = & \dps{ 
          \frac{{\tilde \xi}_H}{\sqrt{2}}
            ( {\tilde \Pi}_x  -  \i  {\tilde \Pi}_y ) 
            }\\[10pt]
     \tilde \eta^{\dagger} 
          & = & \dps{ 
          \frac{{\tilde \xi}_H}{\sqrt{2}}
            ( {\tilde \Pi}_x  +  \i  {\tilde \Pi}_y ) 
            }
     \end{array}
     }
with 
\Equation{eq:xiH}
{
     \tilde \xi_H = \sqrt{\frac{c}{2|e|{\tilde H}}} 
       = \Bigl [ \frac{m_xm_y}{m_z^2} \Bigr ]^{\frac{1}{12}} \xi_H , 
     }
where $\xi_H = \sqrt{c/2 |e| H} = \sqrt{\Phi_0/2 \pi H}$, 
which is of the order of the BCS coherence length $\xi_0$ when $H \sim H_{c2}$. 
The operator ${\tilde \vv'_F} \cdot {\tilde \vPi}$, 
which appears in Eq.~\refeq{eq:Hc2eq}, can be rewritten as 
\Equation{eq:vFPitilde}
{
     \begin{array}{rcl} 
     {\tilde \vv'_F} \cdot {\tilde \vPi} 
     & = & \dps{ 
     \frac{1}{ \sqrt{2} {\tilde \xi}_H }
     {\tilde v}_F 
     \sin {\tilde \theta}'
          (   \e^{  \i {\tilde \varphi}' } {\tilde \eta}
            + \e^{- \i {\tilde \varphi}' } {\tilde \eta}^{\dagger} ) 
     }\\ [10pt]
     & & \dps{ 
     \hsp{4} 
       - \i \hbar 
         {\tilde v}_F \cos {\tilde \theta}'
          \frac{\partial}{\partial {\tilde z}} 
     } , 
     \end{array}
     }
where 
${\tilde \theta}'$ and ${\tilde \varphi}'$ denote 
the polar coordinates when ${\tilde z}$-axis is the polar axis.

Because \eq.{eq:Hc2eq} can be regarded as an eigen equation with 
the eigenvalue $- \log \bigl ( {T}/{T_c^{(0)}} \bigr )$ 
and eigenfunction $\Delta_{\alpha}(\vr)$, 
our problem is reduced to finding the eigenfunctions 
with the highest eigenvalue. 
The solutions can be written in the form 
\Equation{eq:Deltaphiez}
{
     \Delta(\vr) 
       = {\bar \Delta}({\tilde x},{\tilde y}) 
                       \, \exp[ \i {\tilde q}_z {\tilde z} / \hbar] , 
     }
with $\partial/\partial {\tilde z}$ in Eq.~\refeq{eq:vFPitilde} 
replaced by $\i {\tilde q}_z/\hbar$. 
The gap equation~\refeq{eq:Hc2eq} can be rewritten as 
\Equation{eq:Hc2eqqz}
{
     \begin{array}{rcl}
     \lefteqn{ 
     \dps{ 
     - \log \bigl ( \frac{T}{T_c^{(0)}} \bigr ) 
     \phi({\tilde x},{\tilde y}) 
     } } \\[12pt]
     & = & 
     \dps{ 
     \pi T \int_0^{\infty} \d t \frac{1}{\sinh( \pi T t )} 
          \int \frac{\d \Omega_{{\tilde \vp}'}}{4 \pi} 
          \bigl [ \gamma_{\alpha}( {\hat \vp}' ) \bigr ]^2 
         } \\[12pt]
     && 
     \dps{ 
          \times 
          \Bigl [ 1 
          - \cos \bigl [  
               t \bigl ( 
               h - \frac{1}{2}
                        {\tilde v}_F 
                   {\tilde q}_z \cos {\tilde \theta}' 
                 - {\hat \zeta} 
                 \bigr ) 
                 \bigr ]  
          \Bigr ] 
          \phi({\tilde x},{\tilde y}) , 
            }
     \end{array}
     }
where we have defined 
\Equation{eq:zetadef}
{
     {\hat \zeta} = 
                \frac{{\tilde v}_F \sin {\tilde \theta}' }
                     {2\sqrt{2}{\tilde \xi}_H} 
                (   \e^{   \i {\tilde \varphi}'} {\tilde \eta} 
                  + \e^{ - \i {\tilde \varphi}'} {\tilde \eta}^{\dagger} ) . 
     }

It is convenient to expand 
the function $\phi({\tilde x}, {\tilde y})$ 
by the Abrikosov functions $\phi_n^{(k)}({\tilde x}, {\tilde y})$ 
defined by 
\Equation{eq:Abrikosov}
{
     \phi_n^{(k)}({\tilde x}, {\tilde y}) 
       = \frac{1}{\sqrt{n!}} ({\tilde \eta}^{\dagger})^{n} 
            \phi_0^{(k)} ({\tilde x}, {\tilde y}) , 
     }
where $n = 0,1,2,3,\cdots$, called the Landau level indexes, 
$k$ is an arbitrary wave number, 
and $\phi_0^{(k)}$ is the solution of 
\Equation{eq:phi0}
{
     {\tilde \eta} \, \phi_0^{(k)} ({\tilde x}, {\tilde y}) = 0 , 
     }
which is expressed as 
\Equation{eq:phi0expression}
{
     \phi_0^{(k)} ({\tilde x}, {\tilde y}) = 
     C \e^{\i k {\tilde x}} 
           \exp \Bigl[ 
                      - \frac{( {\tilde y} - {\tilde y}_k)^2} 
                             {2 {\tilde \xi_H}^2} 
                      \Bigr ] , 
     }
with ${\tilde y}_k \equiv k {\tilde \xi}_H^2$ 
and a normalization constant $C$. 
The function $\phi_n^{(k)}({\tilde x}, {\tilde y})$ is expressed as 
\Equation{eq:Hermitepol}
{
     \begin{array}{rcl} 
     \phi_n^{(k)}({\tilde x}, {\tilde y}) 
     & = & \dps{ 
     (-1)^n C \e^{\i k {\tilde x}} 
       H_n \Bigl[ 
       \sqrt{2}
       \frac{ {\tilde y} - {\tilde y}_k } 
            {{\tilde \xi}_H} 
       \Bigr ] }\\[10pt]
     && \dps{ 
     \times \exp \Bigl[ 
                      - \frac{( {\tilde y} - {\tilde y}_k)^2} 
                             {2 {\tilde \xi_H}^2} 
                      \Bigr ] . 
     }
     \end{array}
     }
in terms of the Hermite polynomial. 
The operators ${\tilde \eta}$ and ${\tilde \eta}^{\dagger}$ and 
the Abrikosov functions $\phi_n^{(k)}$ satisfy the relations 
\Equation{eq:etaphi}
{
     \begin{array}{rcl} 
     {\tilde \eta}^{\dagger} \, \phi_n^{(k)} ({\tilde x}, {\tilde y}) 
       & = & 
       \sqrt{n + 1} \,\, \phi_{n+1}^{(k)} ({\tilde x}, {\tilde y}) \\[4pt]
     {\tilde \eta} \, \phi_n^{(k)} ({\tilde x}, {\tilde y}) 
       & = & 
       \sqrt{n} \,\, \phi_{n-1}^{(k)} ({\tilde x}, {\tilde y}) . \\
     \end{array}
     }
If we expand the eigenfunctions as 
\Equation{eq:Deltaexpandbyphin}
{
     \Delta(\vr) = \sum_{n = 0}^{\infty} 
                   \Delta_n \phi_n^{(k)} ({\tilde x}, {\tilde y}) 
                       \, \exp[ \i {\tilde q}_z {\tilde z} / \hbar] , 
     }
the gap equation~\refeq{eq:Hc2eq} can be written as a matrix equation 
for the eigenvector 
with the vector elements $\Delta_0,\Delta_1,\Delta_2,\cdots$, as 
\Equation{eq:gapeqDeltan}
{
     - \log \bigl ( \frac{T}{T_c^{(0)}} \bigr ) \Delta_n 
     = \sum_{n'} D_{nn'} \Delta_{n'} 
     }
with 
\Equation{eq:Dnndef}
{
     \begin{array}{rcl} 
     \lefteqn{ \dps{ 
     D_{nn'} 
     = 
          \pi T \int_0^{\infty} \d t \frac{1}{\sinh( \pi T t )} 
         }} \\[12pt]
     && 
     \dps{ 
          \times 
             \int \frac{\d \Omega_{{\tilde \vp}'}}{4 \pi} 
               \bigl [ \gamma_{\alpha}( {\hat \vp}' ) \bigr ]^2 
          \int \d {\tilde x} \, \d {\tilde y} \,\, 
          \phi_{n}^{(k)}({\tilde x},{\tilde y}) 
         } \\[12pt]
     && 
     \dps{ 
          \times 
          \Bigl [ 1 
          - \cos \bigl [  
               t \bigl ( 
               h 
               - \frac{1}{2}
                        {\tilde v}_F 
                   {\tilde q}_z \cos {\tilde \theta}' 
                 - {\hat \zeta} 
                 \bigr ) 
                 \bigr ]  
          \Bigr ] 
          \phi_{n'}^{(k)}({\tilde x},{\tilde y}) . 
            }
     \end{array}
     }
When $h = 0$, for $s$-wave pairing, 
the solution with $n = 0$ and $q_z = 0$ gives 
the highest upper critical field. 
In general, Abrikosov functions with different $n$'s can be mixed.

The magnetic field $|\vH|$ appears both 
in the Hamiltonians $H_0$ and $H_m$. 
The field $|\vH|$ which originates from $\vA$ in $H_0$ 
is responsible for the orbital effect, 
and 
appears in the gap equation as a dimensionless parameter $a_m |\vH|$ 
with the coefficient $a_m$ defined by 
\Equation{eq:amtildedef}
{
     a_m \equiv \frac{\tilde m}{\sqrt{m_xm_y}}
          \frac{2|e|}{c} \Bigl ( \frac{v_F}{2\pi T_c^{(0)}} \Bigr )^2 . 
     }
In contrast, 
the field $|\vH|$ included in the Zeeman field $h$ in $H_m$ 
is responsible for the Pauli paramagnetic pair-breaking effect. 
For this field, it is convenient to define the dimensionless parameter 
$\mue |\vH|/(2\pi T_c^{(0)})$. 
The relative strength of the Pauli paramagnetic pair-breaking effect 
to that of the orbital effect 
is expressed by the ratio of their dimensionless parameters 
\Equation{eq:zmdef}
{
     z_m = \frac{\mue |\vH|/(2\pi T_c^{(0)})}{a_m |\vH|}
     = \frac{\mue}{2 \pi T_c^{(0)} a_m} . 
     }
The parameter $z_m$ is proportional to the Maki parameter $\alpha$. 
For example, for $s$-wave pairing, 
numerical calculations give $a_m H_{{\rm c}20} \approx 1.0372$ 
and $\mue H_{\rm P}/\Delta_0 \approx 0.707107$, 
and hence $\alpha \approx 7.39 \times z_m$.

If we define ${\bar a}_{m}$ and ${\bar z}_{m}$ for the isotropic system by 
\Equation{eq:amzmdef}
{
     \begin{array}{rcl}
     {\bar a}_{m} & \equiv &  (2|e|/c) (v_F/2\pi T_c^{(0)})^2  \\
     {\bar z}_{m} & =      &  \mue/(2 \pi T_c^{(0)} {\bar a}_{m}) , 
     \end{array}
     }
we obtain 
\Equation{eq:alpha_alphaiso}
{
     z_m = \Bigr ( \frac{m_xm_y}{m_z^2} \Bigr )^{1/6} {\bar z}_{m} , 
     \hsp{3}
     \alpha
     = \Bigr ( \frac{m_xm_y}{m_z^2} \Bigr )^{1/6} {\bar \alpha} , 
     }
with the Maki parameter of the isotropic system 
${\bar \alpha} \equiv \sqrt{2} {\tilde H}_{{\rm c}20}/H_{\rm P}$. 
For the magnetic field parallel to $z$-axis, 
when $m_x \gg m_z$ we obtain $z_m \gg {\bar z}_{m}$, {\it i.e.}, 
$\alpha \gg {\bar \alpha}$, 
which means that the system is strongly Pauli paramagnetic limited.

\section{\label{sec:crossoverrelation}
crossover from the pure FFLO state 
to the coexistence state 
}

Now, we consider a quasi-two-dimensional system such that 
$m_x \gg m_y, m_z$. 
In this case, 
because we have ${\tilde \xi_H} \gg \xi_H \sim \xi_0 $ from \eq.{eq:xiH}, 
we may omit 
${\hat \zeta} \sim 
(   \e^{   \i {\tilde \varphi}'} {\tilde \eta} 
  + \e^{ - \i {\tilde \varphi}'} {\tilde \eta}^{\dagger} )/{\tilde \xi}_H$ 
in Eq.~\refeq{eq:Dnndef} for finite $n$ and $n'$. 
Therefore, if we can truncate the summation over $n$ 
in \eq.{eq:Deltaexpandbyphin}, 
the upper critical field equation is reduced to 
\Equation{eq:Hc22Dlimit}
{
     \begin{array}{rcl}
     \lefteqn{ \dps{ 
     - \log \bigl ( \frac{T}{T_c^{(0)}} \bigr ) 
     = 
     \pi T \int_0^{\infty} \d t \frac{1}{\sinh( \pi T t )} 
       } } \\[12pt]
     && 
     \dps{ 
          \times 
          \int \frac{\d \Omega_{{\tilde \vp}'}}{4 \pi} 
          \bigl [ \gamma_{\alpha}( {\hat \vp}' ) \bigr ]^2 
          \Bigl [ 
           1 - \cos \bigl [ t \bigl ( 
               h - \frac{1}{2} {\tilde v}_F {\tilde q}_z \cos {\tilde \theta}' 
                    \bigr ) \bigr ] 
          \Bigr ] , 
            }
     \end{array}
     }
in the limit ${\tilde \xi}_H \rightarrow \infty$, 
which coincides with the upper critical field equation 
for the pure FFLO state expressed by Eq.~\refeq{eq:FFstate} 
with ${\tilde \vq} \equiv (0,0,{\tilde q}_z)$. 
The magnitude $|\tilde \vq| = |{\tilde q}_z|$ should be optimized 
so that the upper critical field is maximized. 
We write the optimum value as $q_0$. 
For example, for $s$-wave pairing, it is known that 
$q_0 \approx 1.2 \times 2h/{\tilde v}_F$ at $T = 0$. 
Thus, we only have two states with the highest upper critical field, 
which have the FFLO vectors ${\tilde \vq} = (0,0, \pm q_0)$. 
This contradicts the fact that 
there are more than two $\e^{\i {\tilde \vq}_m \cdot {\tilde \vr}/\hbar} $ 
in most exactly two-dimensional systems 
as mentioned below Eq.~\refeq{eq:FFstate}. 
Furthermore, below the upper critical field, 
the free energy is minimized by the state expressed 
by the linear combination 
of more than two $\e^{\i {\tilde \vq}_m \cdot {\tilde \vr}/\hbar}$ 
at low \mbox{temperatures~\cite{Shi98a,Bow02,Mor05}}.

This contradiction is due to the assumption that 
infinitely large $n$'s are negligible in the two-dimensional limit. 
In the limit $m_x \gg m_y, m_z$, we need to consider 
Abrikosov functions with infinitely large Landau level index $n$. 
For large $n$'s, 
the states with $n \pm 1$ can be approximated by the state with $n$, 
which means that $\phi_{n \pm 1} \approx \e^{\pm \i \varphi_0} \phi_n$ 
in Eq.~\refeq{eq:etaphi}, 
where $\e^{ \pm \i \varphi_0}$ are arbitrary phase factors. 
Therefore, we may write 
\Equation{eq:etalimit}
{
     {\tilde \eta} = \sqrt{n} \e^{- \i \varphi_0} , 
     \hsp{4}
     {\tilde \eta}^{\dagger} = \sqrt{n} \e^{\i \varphi_0} 
     }
in the gap equation. 
This procedure is analogous to 
those in the theory of Bose condensation, 
and in two-dimensional type-II superconductors 
in a tilted magnetic field~\cite{Shi97a}. 
Using Eq.~\refeq{eq:etalimit} we obtain 
\Equation{eq:zetalargen}
{
     {\hat \zeta} = 
                \frac{ \sqrt{n} {\tilde v}_F \sin {\tilde \theta}' }
                     {\sqrt{2}{\tilde \xi}_H} 
                \cos ({\tilde \varphi}' - \varphi_0) . 
     }
Thus, the gap equation Eq.~\refeq{eq:Hc2eqqz} can be rewritten 
for $m_x \gg m_y, m_z$ as 
\Equation{eq:Hc2correctlimit}
{
     \begin{array}{rcl}
     \lefteqn{ 
     \dps{ 
     - \log \bigl ( \frac{T}{T_c^{(0)}} \bigr ) 
     \phi({\tilde x},{\tilde y}) 
     } } \\[12pt]
     & = & 
     \dps{ 
     \pi T \int_0^{\infty} \d t \frac{1}{\sinh( \pi T t )} 
          \int \frac{\d \Omega_{{\tilde \vp}'}}{4 \pi} 
          \bigl [ \gamma_{\alpha}( {\hat \vp}' ) \bigr ]^2 
         } \\[12pt]
     && 
     \dps{ 
          \times 
          \Bigl [ 1 
          - \cos \bigl [  
               t \bigl ( 
               h - \frac{1}{2} {\tilde \vv}_F' \cdot {\tilde \vq} 
                 \bigr ) 
                 \bigr ]  
          \Bigr ] 
          \phi({\tilde x},{\tilde y}) , 
            }
     \end{array}
     }
where 
\Equation{eq:qtildedef}
{
     {\tilde \vq} = ({\tilde q}_x, {\tilde q}_y, {\tilde q}_z ) 
                  = ({\tilde q}_{\perp} \cos \varphi_0, 
                     {\tilde q}_{\perp} \sin \varphi_0, 
                     {\tilde q}_z ) 
     }
with 
\Equation{eq:qperpdef}
{
     {\tilde q}_{\perp} = \sqrt{ {\tilde q}_x^2 + {\tilde q}_y^2 } 
                        = \frac{\sqrt{2 n}}{{\tilde \xi}_H}
                        = \sqrt{\frac{2 n {\bar \alpha}}{\alpha}}
                          \frac{1}{\xi_H} . 
     }
If we write the optimum $n$ for each fixed $\alpha$ as $n(\alpha)$, 
we obtain 
\Equation{eq:qperpdeflimit}
{
     {\tilde q}_{\perp} 
     = \lim_{\alpha \rightarrow \infty} 
         \sqrt{\frac{2 n(\alpha) {\bar \alpha}}{\alpha \xi_H^2}} , 
     }
If ${\tilde q}_{\perp} \ne 0$ in the limit $\alpha \rightarrow \infty$, 
$n(\alpha)$ must diverge like $\alpha$, {\it i.e.}, $n(\alpha) \sim \alpha$. 
Equation~\refeq{eq:Hc2correctlimit} coincides with the upper critical 
field equation of the pure FFLO state with ${\tilde \vq}$.

Equation~\refeq{eq:qperpdeflimit} is an essential equation which 
connects the pure FFLO state with the optimum $\vq$ 
in the two-dimensional limit 
and the coexistence state with the optimum $n$ and ${\tilde q}_z$ 
in quasi-two-dimensions. 
From \eqs.{eq:qperpdef} and \refeq{eq:qperpdeflimit}, 
we can see that if there are more than two optimum ${\tilde \vq}$'s 
in the two-dimensional limit, 
there must be coexistence states with different $n$'s 
with close upper critical fields in quasi-two-dimensions 
where $m_x \gg m_y, m_z$.

For example, a $d$-wave superconductor with 
\Equation{eq:gammad}
{
     \gamma_{d_{y^2-z^2}}({\hat \vp}) 
          \propto {\hat p}_y^2 - {\hat p}_z^2 
     }
at low temperatures 
exhibits a degeneracy in the solutions of Eq.~\refeq{eq:Hc2correctlimit} 
with ${\tilde \vq} = (0,\pm q_0,0)$, $(0,0,\pm q_0)$ 
in the limit $m_x \rightarrow \infty$. 
From Eqs.~\refeq{eq:qtildedef} and \refeq{eq:qperpdeflimit}, 
we find that when $m_x \gg m_y,m_z$, 
the coexistence states 
$\Delta(\vr) \approx \Delta_n \phi_{n}^{(k)}({\tilde x},{\tilde y}) 
\e^{\i {\tilde \vq}\cdot {\tilde \vr}/\hbar}$ 
with $(n,{\tilde q}_z) \approx (q_0^2 {\tilde \xi_H}^2 /2,0)$ 
and those with $(n,{\tilde q}_z) = (0,\pm q_0)$ 
have upper critical fields close to each other.

\section{\label{sec:swavepairing}
$s$-wave pairing
}

In this section, we consider an $s$-wave superconductor 
as an example. 
Since $\gamma_{s}(\vp) = 1$, 
the gap equation~\refeq{eq:Hc2eq} 
is exactly the same as that of the isotropic model except that 
the vector potential $\vA$ is scaled as Eq.~\refeq{eq:H0scaled} 
and the eigenfunctions are distorted. 
Because the Zeeman field $h$ is not scaled, 
in contrast to the vector potential $\vA$, 
the Maki parameter $\alpha$ changes from ${\bar \alpha}$ as expressed 
in \eq.{eq:alpha_alphaiso}.

We expand the gap equation \refeq{eq:Hc2eqqz} 
with respect to the operators ${\tilde \eta}$, 
and obtain the eigenfunctions 
\Equation{eq:Deltaswave} 
{
     \Delta(\vr) = \Delta_n \phi_n^{(k)} ({\tilde x}, {\tilde y}) 
                       \, \exp[ \i {\tilde q}_z {\tilde z} / \hbar] , 
     }
which are indexed by $n$ and ${\tilde q}_z$. 
The upper critical field equation is decoupled into 
those for each eigenfunction as 
\Equation{eq:swavegapeq} 
{
     \begin{array}{rcl}
     \lefteqn{ 
     \dps{ 
     - \log \bigl ( \frac{T}{T_c^{(0)}} \bigr ) 
     = 
     \pi T \int_0^{\infty} \d t \frac{1}{\sinh( \pi T t )} 
          \int_0^{\pi/2} \sin \theta \d \theta 
     } } \\[12pt]
     && 
     \dps{ 
          \times 
          \Bigl [ 1 
          - \cos (ht) 
            \cos \bigl [ \frac{1}{2} {\tilde q}_z {\tilde v}_F t 
                         \cos \theta \bigr ] 
          \exp \bigl [ - \frac{{\tilde v}_F^2}{16 {\tilde \xi}_H^2}
                         t^2 \sin^2 \theta \bigr ]
         } \\
     && 
     \dps{ 
          \hsp{4} \times 
          \sum_{m=0}^{n} (-1)^m 
            \bigl [ \frac{1}{8 {\tilde \xi}_H^2} {\tilde v}_F^2 t^2 
                    \cos^2 \theta \bigr ]^m 
          \frac{n!}{(m!)^2 (n-m)!} 
          \Bigr ] . 
         } \\
     \end{array}
     }
The physical upper critical field $H_{\rm c2}(T)$ is 
the highest solution of $H$ among the solutions of \eq.{eq:swavegapeq} 
at each fixed $T$. 
In other words, the parameters $n$ and ${\tilde q}_z$ are optimized 
so that $H_{\rm c2}$ is maximized.

In systems with $\vA = {\vector 0}$, 
the pure FFLO state occurs as described in section~\ref{sec:pureFFLOstate} 
at low temperatures. 
When the system is isotropic, 
there is infinite degeneracy with respect to the direction of $\vq$. 
We express the $\vq$'s that give the maximum FFLO critical field as 
$\vq = (q_0 \sin \theta \cos \varphi, q_0 \sin \theta 
     \sin \varphi, q_0 \cos \theta)$ 
with an optimum value of $q_0$ and arbitrary $\theta$ and $\varphi$.

In an anisotropic system with $m_x \gg m_y, m_z$, 
the effective Maki parameter becomes large so that 
$\alpha \gg {\bar \alpha}$ from \eq.{eq:alpha_alphaiso}, 
and ${\tilde \xi}_H \gg \xi_H$ from \eq.{eq:xiH}. 
When $m_x$ is very large, 
we can make approximate $|\vq| \approx q_0$. 
Therefore, from Eqs.~\refeq{eq:qtildedef} and \refeq{eq:qperpdef}, 
we obtain the optimum value of ${\tilde q}_z$, as 
\Equation{eq:qzswave}
{
     {\tilde q}_z = \pm \sqrt{ q_0^2 - \frac{2n}{{\tilde \xi}_H^2} } 
     = \pm \sqrt{ q_0^2 - \frac{2n {\bar \alpha}}{\alpha \xi_H^2} } 
     }
for a given $n$. 
If we consider that $n$ is always finite 
for the limit $\alpha \rightarrow \infty$, 
\eq.{eq:qzswave} is reduced to ${\tilde q}_z = \pm q_0$, 
and the $H_{\rm c2}$ equation \refeq{eq:swavegapeq} 
is reduced to that of $n = 0$. 
Therefore, when $m_x$ is very large, states with any finite $n$ have 
upper critical fields very close to 
that of the state with $n = 0$ and ${\tilde q}_z = \pm q_0$. 
However, because the index $n$ needs to be optimized for each situation, 
it can be infinitely large. 
For the state with infinitely large $n$, 
such that $n \propto \alpha \propto {\tilde \xi}_H^2$, 
\eq.{eq:qzswave} results in $|{\tilde q}_z| < q_0$ 
and ${\tilde q}_{\perp} \ne 0$.

\begin{figure}[htbp]
\vspace{4ex} 
\includegraphics[width=6.5cm]{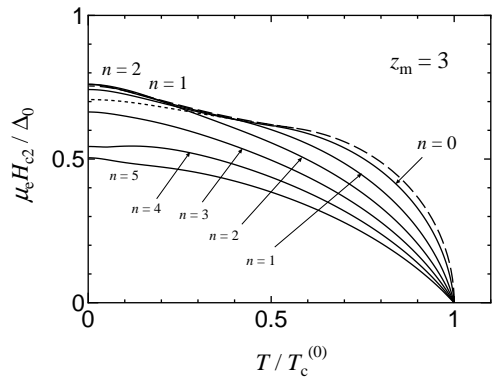}
\caption{
Temperature dependences of the upper critical fields 
for $z_m = 3$. 
The solid curves show the upper critical fields for $n=0,1,2,3,4,5$. 
At each temperature, the highest field is the physical result of 
the critical field. 
The broken and dotted curves show the upper critical fields 
in the two-dimensional limit 
and that with the assumption of $\vq = {\vector 0}$. 
} 
\label{fig:Hc2zm3}
\end{figure}

This can be verified also by numerical calculations. 
Figure~\ref{fig:Hc2zm3} shows the temperature dependences of the 
critical fields of states of various $n$. 
At each temperature, 
the state with the highest critical field is physical. 
It is found that vortex states with $n=0,1,2$ occur 
depending on the temperature, 
and that the envelope is very close to 
the curve of the two-dimensional limit. 
It is also found that the wave vector $\vq$ becomes nonzero 
below $T \approx 0.51 \times T_c^{(0)}$.

\begin{figure}[htbp]
\vspace{4ex} 
\includegraphics[width=6.5cm]{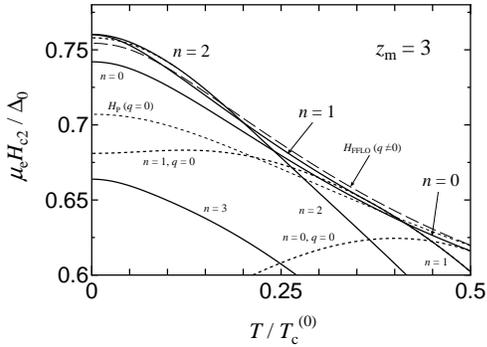}
\caption{
Temperature dependences of the upper critical fields 
for $z_m = 3$ at low temperature. 
The solid curves show the upper critical fields for $n=0,1,2,3$. 
At each temperature, the highest field is the physical result of 
the critical field. 
The dotted curves show the critical field 
when $\vq = {\vector 0}$ is assumed. 
The thin broken and dotted curves show the upper critical field 
in the two-dimensional limit 
and that with the assumption of $\vq = {\vector 0}$. 
} 
\label{fig:Hc2zm3LT}
\end{figure}

Figure~\ref{fig:Hc2zm3LT} shows the behavior of the critical field 
for each $n$ at low temperatures. 
We obtain $q_z \ne 0$ 
below the temperature where the solid and dotted curves branch off. 
Interestingly, below $T \approx 0.18 \times T_c^{(0)}$, 
the upper critical field of 
the coexistence state with $n=2$ and ${\tilde q}_z \ne 0$ 
exceeds that in the limit $z_m \rightarrow \infty$. 
We will discuss this later.

\begin{figure}[htbp]
\vspace{4ex} 
\includegraphics[width=6.5cm]{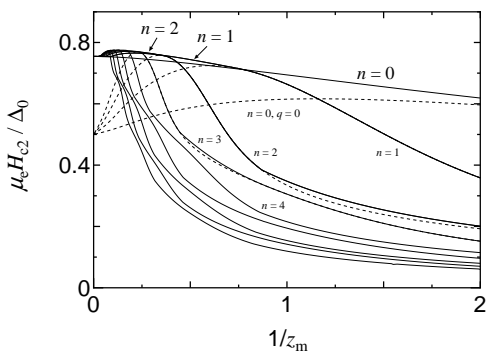}
\caption{
$z_m$ dependences of the upper critical fields for $n=0,1,2,\cdots 8$ 
at $T/T_{\rm c} = 0.01$. 
The dotted curves show those when $\vq = {\vector 0}$ is assumed. 
$\alpha \approx 7.39 \times z_m$. 
} 
\label{fig:Hc2zmdep}
\end{figure}

Figure~\ref{fig:Hc2zmdep} shows 
the $z_m$ ($\propto \alpha$) dependences of 
the upper critical fields $H_{\rm c2}$ 
for $n=0,1,\cdots 8$. 
It is found that the upper critical fields of 
all the coexistence states with $n \ne 0$ tend to 
approach that of the coexistence state with $n=0$, 
and slightly exceed it, where $z_m$ is large. 
At each $z_m$, the highest upper critical field among those with 
$n = 0,1,2,\cdots$ is the physical upper critical field. 
It is found that $n$ of the physical state increases as $z_m$ increases, 
and the physical critical field is larger than 
the critical field of the $n=0$ state for $z_m \gsim 1.3$.

\begin{figure}[htbp]
\vspace{4ex} 
\includegraphics[width=6.5cm]{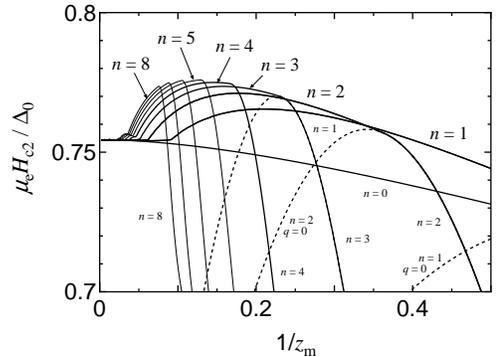}
\caption{
$z_m$ dependences of the upper critical fields for $n=0,1,2,\cdots 8$ 
at $T/T_{\rm c} = 0.01$ in the large $z_m$ region. 
The dotted curves show those when $\vq = {\vector 0}$ is assumed. 
The thick and thin curves show the results for 
$n=0,1,2,3,4$ and those for $n=5,6,7,8$, respectively. 
$\alpha \approx 7.39 \times z_m$. 
} 
\label{fig:Hc2zmdeplargezm}
\end{figure}

Figure~\ref{fig:Hc2zmdeplargezm} shows the behavior of $H_{\rm c2}$ 
for large $z_m$. 
After reaching maxima, 
all the critial fields for $n \ne 0$ decrease, 
and converge with the curve for $n=0$ 
in the limit $z_m \rightarrow \infty$. 
Within the present theory, 
the physical upper critical field, 
{\it i.e.}, the highest field at each $z_m$, 
decreases as $z_m$ increases for very large $z_m \gsim 7.7$, 
{\it i.e.}, $1/z_m \lsim 0.13$. 
This result seems inconsistent with the naive expectation 
that a reduction of the orbital effect, for example, 
by increasing $v_{\rm F}$ should cause a weaker pair-breaking effect. 
Presumably, for such a large $z_m$, 
a first-order phase-transition to a coexistence state must occur 
at a higher critical field than that obtained here, 
and the resultant critical field must monotonically increase with $z_m$.

\section{\label{sec:summary}
summary and discussion
}

We have examined quasi-two-dimensional type-II superconductors 
and the two-dimensional limit. 
When $\vH \parallel [0,0,1]$, 
the effective Maki parameter $\alpha$ is proportional to $(m_xm_y/m_z^2)^{1/6}$. 
Therefore, when $m_x \gg m_z$, 
the orbital pair-breaking effect becomes weak, 
and the superconductivity survives up to a higher field, 
where the FFLO state is favored.

In the coexistence state, the FFLO modulation occurs in the direction of 
the magnetic field, and is smoothly reduced to that of the pure FFLO state 
in the two-dimensional limit ($m_x/m_z \rightarrow 0$), 
when the directions of the magnetic field and $\vq$ of the pure FFLO state 
coincide. 
In contrast, when their directions differ, 
it may appear that the coexistence state is not reduced 
to the pure FFLO state continuously. 
However, in actuality, 
modulation perpendicular to the magnetic field is realized 
in large $n$ vortex states. 
The physical origin of the order parameter modulation 
in vortex states with higher Landau levels 
is the spin polarization energy 
as in the pure FFLO state. 
The coexistence states with optimum $n$ have 
upper critical fields close to that of the pure FFLO state. 
Hence, a cascade transition occurs when $\alpha$ is large, 
analogously to the exactly two-dimensional system 
in a tilted magnetic 
\mbox{field~\cite{Bul73,Buz96,Shi97a,Kle00,Hou00,Hou02,Kle04}}. 
In the two-dimensional limit $m_x \rightarrow \infty$, 
the Landau level index $n$ increases as 
$n \propto \alpha \propto m_x^{1/6}$. 
As a result, 
the coexistence state indexed by $(n,q_{\parallel})$ 
(or the pure vortex state indexed by $n$ when $q_{\parallel} = 0$) 
is continuously reduced to the pure FFLO state with 
$\vq = (q_{\parallel},\vq_{\perp})$.

The relations~\refeq{eq:qtildedef} and \refeq{eq:qperpdeflimit} connect 
the coexistence states of Eq.~\refeq{eq:Deltaswave} 
and the pure FFLO state in the two-dimensional limit. 
The pure FFLO state with $q_{\perp} \ne 0$ corresponds to 
the coexistence state with $n \propto \alpha \rightarrow \infty$, 
and the upper critical fields of 
the coexistence states converges to that of the pure FFLO state. 
This behavior has been confirmed also by numerical calculations.

From these results, we can conclude that 
the FFLO state obtained in a theoretical model without orbital effects 
may emerge as the vortex states with higher Landau level indexes 
in real materials where orbital effects are inevitable. 
In particular, the order parameter modulation due to 
the higher Landau level index is a mark of the FFLO modulation 
perpendicular to the magnetic field.



\end{document}